\documentclass[%
 aip,
 amsmath,amssymb,
 reprint,%
]{revtex4-1}

\usepackage{graphicx}
\usepackage{dcolumn}
\usepackage{bm}
\usepackage[english]{babel}

\usepackage[utf8]{inputenc}
\usepackage[T1]{fontenc}
\usepackage{mathptmx}
\usepackage{siunitx}
\begin{document}

\preprint{AIP/123-QED}

\title[Injection of Positrons into a Dense Electron Cloud in a Magnetic Dipole Trap]{Injection of Positrons into a Dense Electron Cloud in a Magnetic Dipole Trap}

\author{M. Singer}
 \affiliation{Technische Universität München, 85748 Garching, Germany}
 
 \author{M. R. Stoneking}
 \affiliation{Lawrence University, Appleton, WI 54911, USA }
 \affiliation{Max Planck Institute for Plasma Physics, 17491 Greifswald, 85748 Garching, Germany}
 
  \author{E.V. Stenson}
 \affiliation{Max Planck Institute for Plasma Physics, 17491 Greifswald, 85748 Garching, Germany}
 \author{A. Deller}
 \affiliation{Max Planck Institute for Plasma Physics, 17491 Greifswald, 85748 Garching, Germany}
 \author{A. Card}
 \affiliation{Max Planck Institute for Plasma Physics, 17491 Greifswald, 85748 Garching, Germany}
 \author{S. Nißl}
 \affiliation{Technische Universität München, 85748 Garching, Germany}
 \affiliation{Max Planck Institute for Plasma Physics, 17491 Greifswald, 85748 Garching, Germany}
 \author{J. Horn-Stanja}
 \affiliation{Max Planck Institute for Plasma Physics, 17491 Greifswald, 85748 Garching, Germany}
 \author{T. Sunn Pedersen}
 \affiliation{Max Planck Institute for Plasma Physics, 17491 Greifswald, 85748 Garching, Germany}
 \affiliation{University of Greifswald, 17489 Greifswald, Germany}
 \author{H. Saitoh}
 \affiliation{The University of Tokyo,  Kashiwa 277-8561, Japan}
 \author{C. Hugenschmidt}
 \affiliation{Technische Universität München, 85748 Garching, Germany}

\date{\today}

\begin{abstract}
The creation of an electron space charge in a dipole magnetic trap and the subsequent injection of positrons has been experimentally demonstrated. Positrons (\SI{5}{\electronvolt}) were magnetically guided from their source and injected into the trapping ﬁeld generated by a permanent magnet (\SI{0.6}{\tesla} at the poles) using a cross ﬁeld E$\times$B drift, requiring tailored electrostatic and magnetic ﬁelds.
The electron cloud is created by thermionic emission from a tungsten filament. The maximum space charge potential of the electron cloud reaches \SI{-42}{\volt}, which is consistent with an average electron density of \num{4 +- 2 e12}\si{\per\cubic\meter} and a Debye length of \SI{2 +-1}{\centi\meter}. We demonstrate that the presence of this space potential does not hamper efficient positron injection. Understanding the effects of the negative space charge on the injection and conﬁnement of positrons represents an important intermediate step towards the production of a conﬁned electron--positron pair plasma.

\end{abstract}

\maketitle

\section{\label{sec:1}Introduction}

Efforts to create pair plasmas consisting of quasi-neutral combinations of particles of opposite charge and equal mass are being pursued across a wide range of regimes \cite{Sarri_2015,Chen_2015,Higaki_2017,Oohara_2003,stoneking_JPP_2020}.  Successful trapping of pair plasma would create a novel and very likely stable system \cite{Tsytovich1978,Iwamoto_1993,helander_connor_2016,Zank_1995}. In particular, production of a low-temperature magnetically well-confined electron-positron plasma would offer numerous opportunities for experimental tests of the theoretical predictions. It is, however, a significant experimental challenge to achieve the density threshold for plasma phenomena in the laboratory with a collection of electrons and positrons. In a traditional quasi-neutral plasma, positive ions and electrons are born together via the ionization of neutral gas.  While one approach to creating electron-positron plasma is to first produce copious quantities of positronium \cite{Deller_2016} which can then be ionized in a manner similar to the ionization of neutral gas, there are significant technical challenges to this method.  For our purposes, a more feasible approach is to assemble a pair plasma by mixing initially separate populations of positive and negative charges.  Established devices for confining either positrons or electrons are called Penning-Malmberg traps, and nested combinations of such devices have been used to bring particles of opposite charge into interaction with one another \cite{Greaves_1995,Gilbert_2001,Hall_1996,Gabrielse_1999,Amole_2013,Kabantsev_2007}.  However, such traps are not suitable for long confinement of quasi-neutral pair plasmas with small Debye lengths. Other efforts have employed non-neutral plasmas or dense beams as electrostatic traps for particles of the opposite sign of charge \cite{oshima2004,Levine_1988}. However, these experiments did not seek to produce quasi-neutral combinations of particles that are of interest here. 

The APEX (A Positron-Electron eXperiment) collaboration aims to confine a magnetized electron-positron plasma in toroidal traps with closed field lines that allow the simultaneous trapping of both species.  Since 2014, we have been working toward achieving this in a levitated dipole trap, a geometry that provides excellent confinement properties for both quasi-neutral and non-neutral plasmas \cite{Yoshida_2012,Baitha_2020,Boxer2010,Saihtoh_PRE_2016}. We have also recently started to pursue development of an optimized stellarator as a second type of confinement device with suitable and complementary properties. Meanwhile, magnetic mirror traps are being explored by Higaki, \emph{et al.}\cite{Higaki_2017}. In the ultimate realization of the dipole branch of our project, positrons from an intense source will be accumulated and delivered in large pulses (> \num{e8}) to a trap consisting of a levitated superconducting coil that is pre-loaded with an electron plasma.  In a series of important preliminary experiments we employed a prototype dipole trap with a supported rare earth magnet to develop the required methods for positron injection, confinement and diagnosis. Important milestones, such as establishing highly efficient positron injection\cite{Stenson_PRL_2018} and long confinement times\cite{Stanja_PRL_2018}, have been achieved. We report here on the next experiments in this series in which we create a dense electron cloud in the prototype trap and measure its impact on our ability to inject positrons.  We find that the presence of an electron cloud with densities and space charge potentials relevant to future experiments does not negatively impact the efficiency of positron injection.  This result represents an important validation of our planned approach to creating electron-positron plasma.

\section{\label{sec:2}Experimental Setup}

One major challenge for establishing a pair plasma is the acces to the requisite number of  positrons. These are provided by the NEPOMUC (NEutron induced POsitron Source MUniCh) positron facility \cite{Hugenschmidt_2012}, situated at the FRM II neutron source. 
The positrons are created in the NEPOMUC beam tube located within the reactor's heavy water moderator tank in close proximity to the fuel element. Thermal neutrons are captured in enriched \si{^{113} Cd}, generating high-energy gamma radiation. These photons subsequently interact with platinum foils to produce electrons and positrons through pair production. These foils are designed as a system of electrostatic lenses that extract the positrons and establish the positron beam. This primary beam provides up to \SI{e9}{\raiseto{+}\elementarycharge\per\second} when set to an accelerating energy of \SI{1}{\kilo\electronvolt} and, beam energies as low as \SI{5}{\electronvolt} are feasible. A narrower spatial and energy distribution is achieved by implanting the  \SI{1}{\kilo\electronvolt} beam into a tungsten single-crystal film and extracting the remoderated positrons \cite{Stanja_NIMA_2016}. For the work described here, we used the remoderated positron beam with an intensity of \num{4.5 +- 0.3 e7} \si{\raiseto{+}\elementarycharge\per\second} and a mean longitudinal energy of \num{4.9 +- 0.7} \si{\electronvolt}.

\begin{figure}[htb]
	\centering
	\begin{minipage} [t]{0.5\textwidth}
		\centering
		\includegraphics[width=0.8\textwidth]{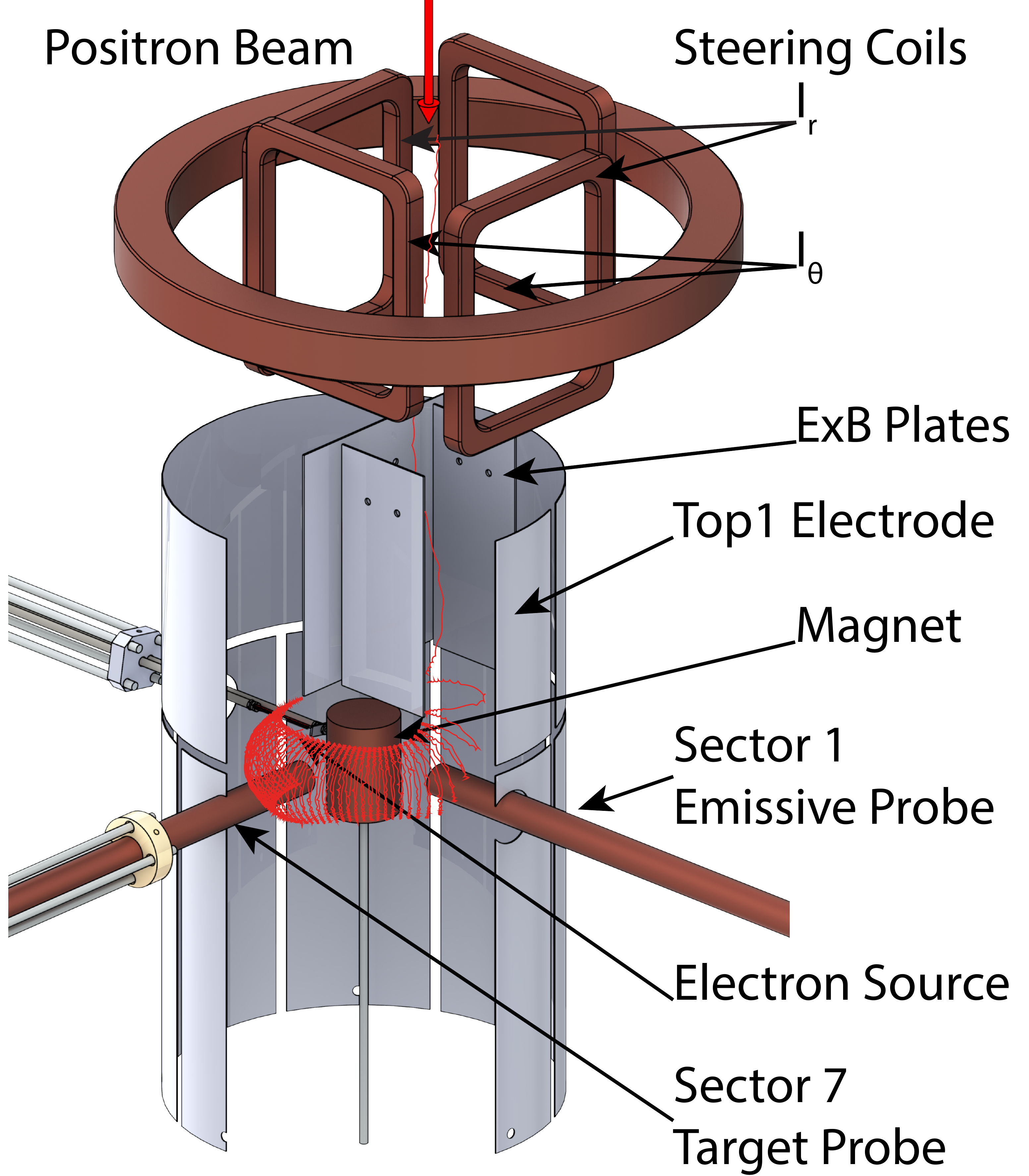}
		\caption{Positron injection into the dipole trap: the positron beam passes the steering coils which are used, to adjust the beam position immediately upstream of the E$\times$B plates. The red line depicts a sample trajectory of a \SI{5}{\electronvolt} positron.}
		\label{fig:chamber_side_view}
	\end{minipage}
\hfill
	\begin{minipage} [t]{0.5\textwidth}
		\centering
	\includegraphics[width=0.8\textwidth]{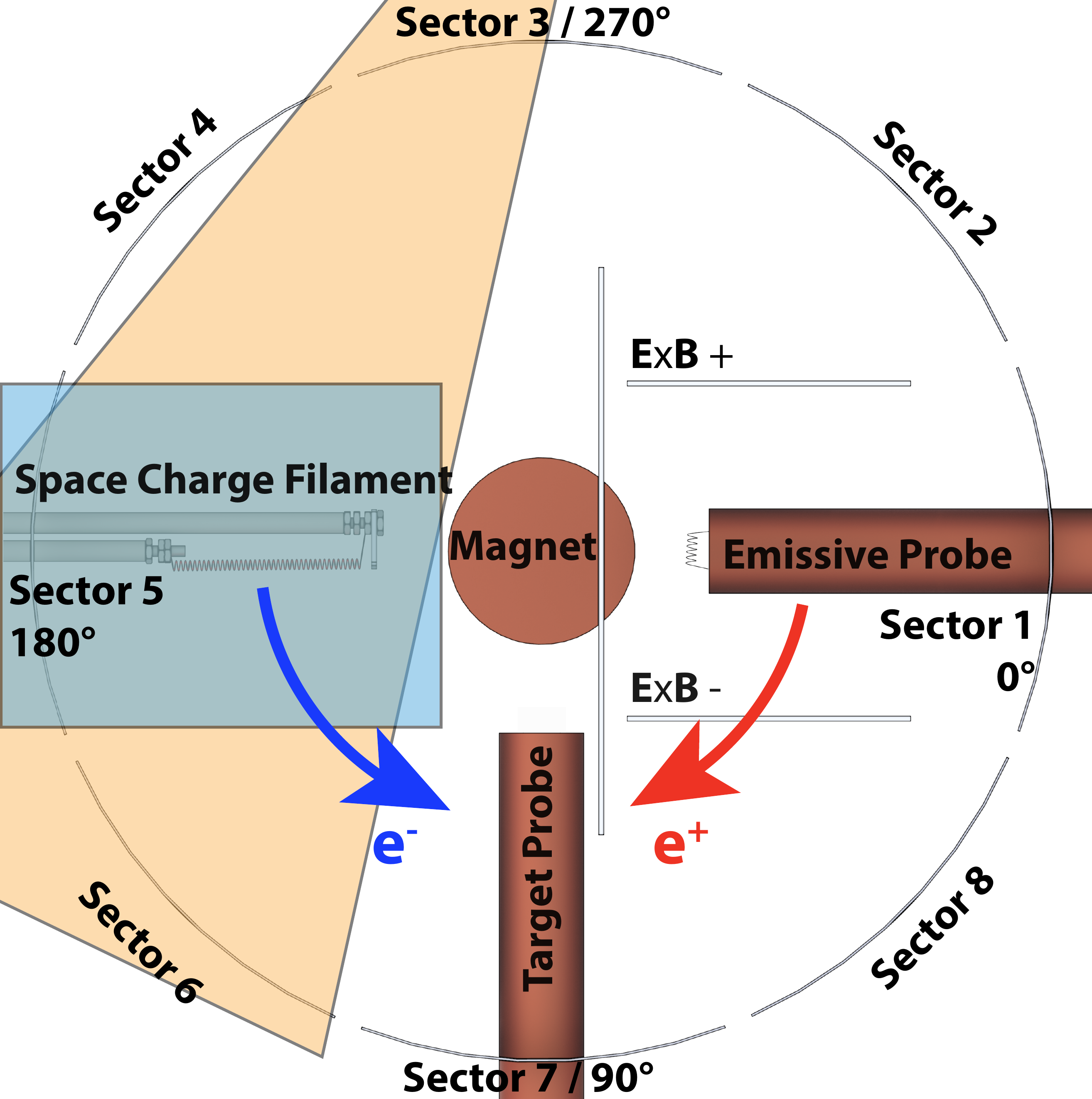}
\caption{Top-down view of the chamber. The yellow triangle represents the field of view of the side BGO detector (A) placed in the equatorial plane of the magnet and the space charge filament assembly. The blue rectangle shows the field of view of the bottom detector (B) in the same plane. The E$\times$B plates are located above the confining area in the region near sector 1 (\ang{0}).}
\label{Fig:chamber_top_view}
	\end{minipage}
	
\end{figure}

The positron beam is adiabatically guided by \SI{5}{\milli\tesla} magnetic fields to the prototype APEX setup \cite{Saitoh_NJP_2015} (Figure \ref{fig:chamber_side_view}). It consists of two vertically stacked vacuum chambers. Immediately upstream of this setup (but not shown here) are a set of diagnostics, including an annihilation target / charge collector, microchannel plate and phosphor screen, and a retarding field analyzer. When the diagnostics are retracted, the beam continues to the experiment setup, passing through two pairs of steering coils.

At the center of the experiment, enclosed in a copper case, is a cylindrical permanent magnet that provides the dipole magnetic field to confined charged particles. The magnet has a diameter of \SI{28}{\milli\meter}, a height of \SI{40}{\milli\meter} and a magnetic field strength of \SI{0.6}{\tesla} at the pole faces. Trapped particles gyrate around the field lines, are magnetically mirrored between the two poles, and drift toroidally around the magnet, due to curvature, $\nabla$ B and E$\times$B drifts. Particles with opposite signs of charge experience opposite toroidal drift directions when the magnetic drifts are dominant. In our configuration, positrons drift clockwise---when viewed from above (Figure \ref{Fig:chamber_top_view})---whereas electrons drift counter-clockwise.

Near the wall of the chamber, electrodes form a cylindrical boundary in two layers. The bottom layer consists of eight equally sized sections (shown in Figure \ref{Fig:chamber_top_view}), while the top layer has a 1/8 (above sector 1 in Figure \ref{Fig:chamber_top_view}) to 7/8 division. In the vicinity of this 1/8 electrode, between the magnet and the  beam line, two oppositely charged, vertically oriented electrodes provide the electric field that generates E$\times$B drift to transport positrons off magnetic field lines that connect to the beam line, onto magnetic field lines that intersect only the permanent magnet (i.e., the confinement region). To minimize the effect of the injection potentials on positrons elsewhere in the trap, a third vertical plate, which is grounded, serves as a shield plate. 

Three instruments can be inserted into the confinement region on its equatorial plane. The positron injection takes place in the \ang{0} area (right side in Figure \ref{Fig:chamber_top_view}). The injected positron beam propagates clockwise. After a \ang{90} drift, a target probe can intercept the positron beam to measure its intensity. At the \ang{180} position, a filament serves both as an annihilation target for positrons and as an electron source. Electrons propagate counterclockwise, therefore the injected electron cloud fills the region through which the injected positrons drift.

The main diagnostics for observing the positron injection are two BGO (bismuth germanate) gamma detectors. Their fields of view are both collimated to monitor annihilation events on the filament assembly. One detector (hereafter referred to as detector A) is placed in the equatorial plane, the other (hereafter referred to as detector B) is placed underneath the chamber. The yellow triangle (Figure \ref{Fig:chamber_top_view}) depicts the approximate line of sight of detector A placed in the equatorial plane of the magnet. The blue rectangle shows the line of sight of detector B placed underneath the chamber.

\section{\label{3}Establishing a Space Charge}

In this section, we describe the methods used to create and characterize the electron cloud into which the positron beam was injected.

\subsection{Electron Source}
\label{sec:electron source}

\begin{figure}[htbp]
	\centering
	\includegraphics[width=0.45\textwidth]{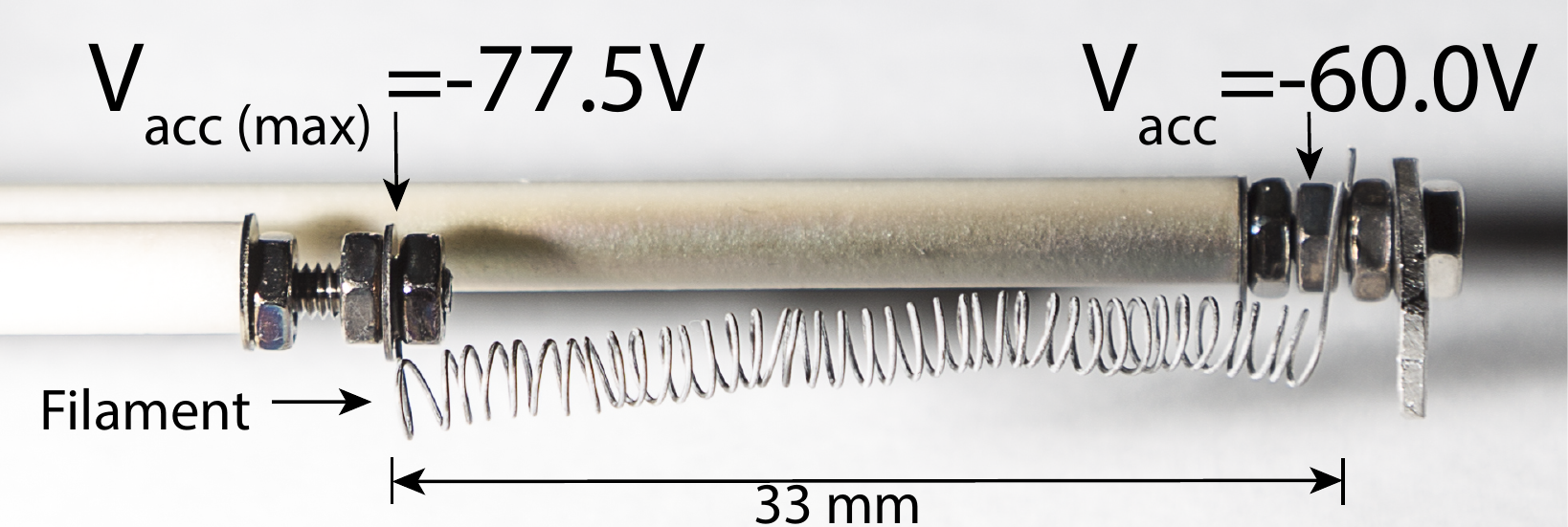}
	\caption{Electron source assembly. Two insulated rods provide the electrical connections to heat and bias the tungsten filament. The plate on the right serves as a heat radiation shield to protect the magnet, which is placed with a distance of \SI{10}{\milli\meter} to its right side. }
	\label{Fig:filament}
	
\end{figure}

The electron source used for this work emits electrons through thermionic emission from a 40--turn coil made of \SI{0.25}{\milli\meter} diameter thoriated tungsten wire with a coil length of \SI{33}{\milli\meter} and an outer coil diameter of \SI{2.8}{\milli\meter} (Figure \ref{Fig:filament}). It is positioned on the equatorial plane of the permanent magnet. The distance between the tip of its assembly and the copper case of the magnet is \SI{10}{\milli\meter}. A current in the range of \si{I_{fil}=} \SIrange{3.4}{4.0}{\ampere} heats the filament, leading to a voltage drop of \si{V_{fil}=} \SIrange{13.5}{17.5}{\volt} across it.
The filament is biased negatively to accelerate electrons away from it (\SI{-60}[V_{acc}=]{\volt}); the end of the filament closest to the magnet is set to \si{V_{acc}}, and the end closest to the wall is typically between \SIrange{-73.5}{-77.5}{\volt}.
The emission current for the configuration with \SI{4.0}[I_{fil}=]{\ampere} is \SI{0.80 +- 0.05}{\milli\ampere}.

Since the filament is placed in the dipole field of the magnet, emitted electrons follow field lines. Electron current measurements on the wall electrodes, the magnet and a target probe inserted at \ang{0} (instead of the emissive probe) indicate that the amount of electrons hitting the wall electrodes is negligible. Approximately 90\% of electrons are immediately lost on the (grounded) magnet. The rest are mirror-trapped and contribute establish a space charge by drifting toroidally (counterclockwise).

\subsection{Measuring the Space Potential}
\label{sec:measuring the space potential}

The space potential of the electron cloud established from the filament source was measured using the emissive probe \cite{Sheehan_2011}, inserted into the equatorial plane at the \ang{0} position. Its working principle can be described using a simple model: when the filament has a more negative potential than the surrounding potential, electrons are emitted. In contrast, when the filament is more positive than the surrounding region, electrons are accelerated towards it and collected. This results in an equilibration of the potentials between filament and its surroundings. If the temperature of the tungsten wire is high enough to enable sufficient emission current, the potential of the electrically floating filament corresponds to the space potential.

\begin{figure}[htbp]
	\centering
	\includegraphics[width=0.49\textwidth]{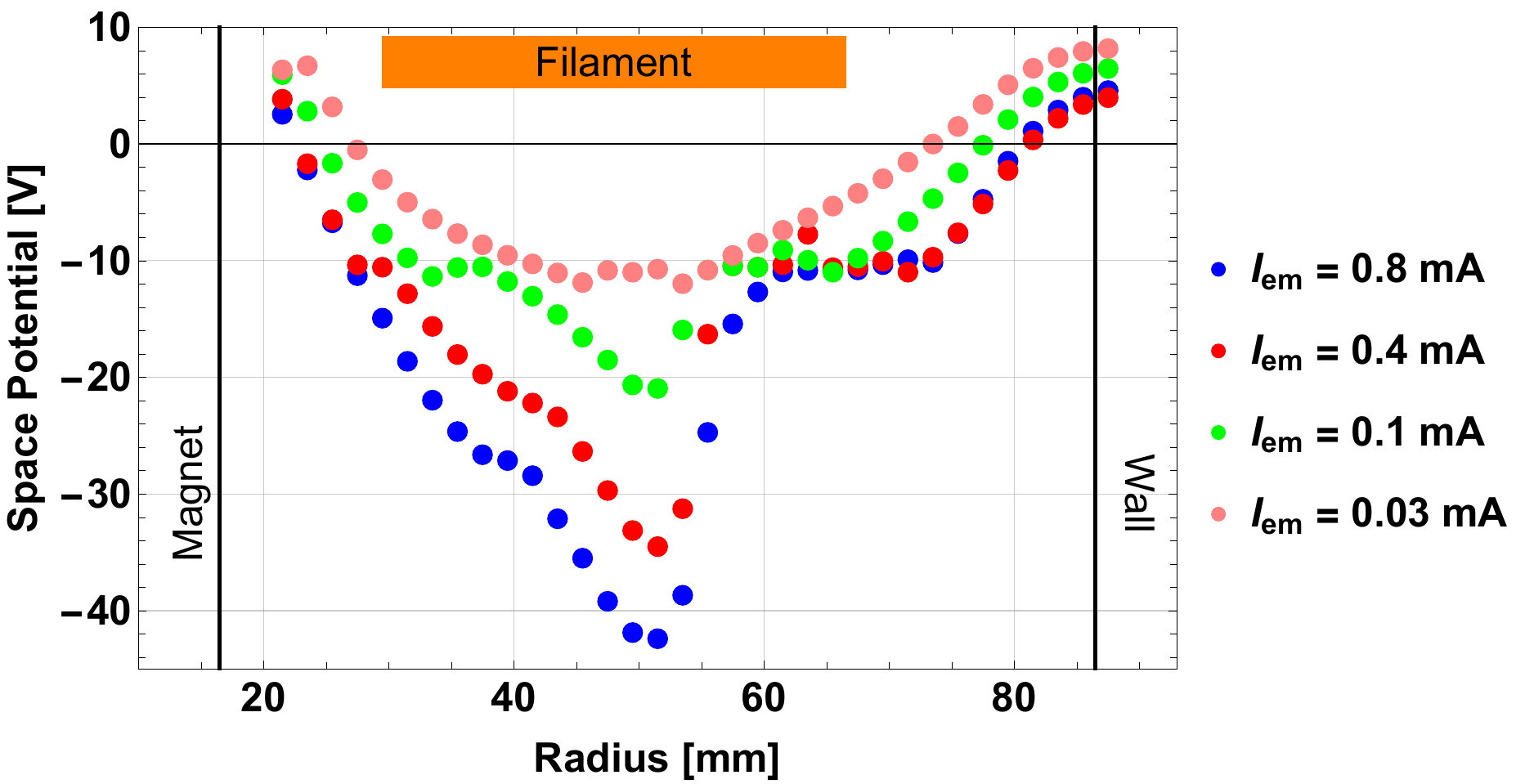}
\caption{Space charge potential profiles for different emission currents, measured with the emissive probe at \ang{0}. The space charge used for the data presented in this work (\SI{0.8}[I_{em}=]{\milli\ampere}), is represented by the blue dots, while the other colors show the space potentials established by lower heating currents. The vertical line at \SI{16}{\milli\meter} shows the radius of the magnet case, while the line at \SI{86}{\milli\meter} shows the radius of the outer wall electrode set. The orange bar illustrates the radial extent of the emitting section of the filament assembly (on the other side of the magnet).}
	\label{Fig:negative_outside}
	
\end{figure}

The supporting probe shaft is covered with insulating Kapton tape and is therefore expected to become negatively charged.  This represents a smaller perturbation to the space potential in the vicinity of the probe tip than a grounded, conducting shaft \cite{Kremer_2007}. The radial profile of the space potential was measured by step-wise retracting the heated filament from the vicinity of the magnet to the outer electrode and measuring the floating potential difference at each step (Figure \ref{Fig:negative_outside}). For all applied heating currents, the most negative potential was measured around position \SI{50}[r=]{\milli\meter}. It can also be seen, that at the ends of the range, in close proximity to the positively biased magnet (\SI{8}[+]{\volt}) and wall electrode (\SI{10}[+]{\volt}), a positive potential was measured. At radii which are populated by electrons, the positive applied potentials are shielded by the electron cloud. This shielding effect also results in nearly identical potential profiles for the configurations with biased (\SI{150}[V_{E \times B}=]{\volt}) and unbiased E$\times$B plates.

There is significant radial structure in the space potential profiles shown in Figure \ref{Fig:negative_outside}, as well as a significant mismatch between the linear potential variation across the filament (\SI{-60}{\volt} at \SI{28}{\milli\meter} to \SI{-77.5}{\volt} at \SI{66}{\milli\meter}).  Without more complete measurement of the three-dimensional structure of the potential we cannot explain this structure in detail.  However, one effect that may be at play is that the portion of the outer wall electrode adjacent to where these measurements took place is biased to \SI{+10}[+]{\volt}.  As the electrons drift into this region they will experience an inward E$\times$B drift that may be responsible for the location of the steepest gradient in the potential occurring closer to the radial location of the middle of the filament rather than its outermost end.

\subsection{Estimating the Electron Density and Debye Length}

We aim to create an electron cloud that is dense enough to constitute a reasonable test of our strategy for creating electron--positron plasma by injection of positrons into a pre-formed pure electron plasma. Additionally, if the electron cloud in these tests is cold enough, it would qualify as a non-neutral plasma if the Debye length is significantly smaller than the system size. The Debye length is defined as:
\begin{equation}
	\lambda _d = \sqrt{\frac{\epsilon _0 k_B T}{n_e e^2}},
	\label{eq:DebyeLength}
\end{equation}
where  \( \epsilon_0\) is the vacuum permittivity, \( k_B\) the Boltzmann constant, \( T\) the electron temperature, \( n_e\) the electron density and \( e\) the elementary charge.

An effective electron temperature can be defined in relation to the mean kinetic energy by:
\begin{equation}
	E_e = \frac{3}{2} k_B T_{\mathrm{eff}} .
	\label{eq:ElectronEnergy}
\end{equation}
A value for the mean kinetic energy is determined from a volume average of the difference between the filament potential (linearly varying from \SIrange{-60}{-77.5}{\volt}) and the space charge potential (Figure \ref{Fig:negative_outside}). For the highest emission current case, this yields a mean energy of \SI{44}{\electronvolt} or an effective temperature of about \SI{30}{\electronvolt}.

There are two (nearly) independent ways we can estimate the order of magnitude of the electron density. First, we observe that 90\% of the emitted electrons are promptly lost on the magnet. The remaining 10\% are mirror trapped and drift toroidally around the magnet. Therefore, for an emission current of \SI{0.80 +- 0.05}[I_{emission}=]{\milli\ampere}, \SI{0.08 +- 0.01}{\milli\ampere} contributes to establishing the space charge. 
Due to the curvature of the magnetic field lines, the shape of the volume occupied by electrons is a torus with a C-shaped cross section. The volume of this torus is determined by modeling this cross section in a CAD program, based on simulated single--particle trajectories. By revolving this shape around the axis of the magnet, a volume of approximately \SI{0.5}{\litre} results. Trajectory calculations \cite{Saitoh_NJP_2015} for \SI{30}{\electronvolt} electrons with an initial radial position of \SI{50}{\milli\metre} indicate that the toroidal drift period is about \SI{3}{\micro\second}. Accordingly, the mean electron density, consistent with these numbers, is \SI{3 +- 1 e12}[n_e=]{\per\cubic\metre}.

The second method for estimating the electron density uses the measured space charge potential and Poisson’s equation, which describes the relation between the electrostatic potential $\varphi$ and charge density.
\begin{equation}
	\nabla^2 \varphi=\frac{n_e e}{\epsilon_0},
	\label{eq:poisson equation}
\end{equation}
To obtain an approximation for the radial profile of the space charge potential, we used several polynomial fits to the data in Figure \ref{Fig:negative_outside} and calculated the Laplacian, assuming cylindrical and up-down symmetry. The volume--averaged density that results is \SI{4 +- 2 e12}{\per\cubic\meter}, which is consistent with the first method.

Importantly, these density estimates are in the target range for planned pair plasma experiments. The estimated temperature, meanwhile, is about an order of magnitude larger than what is ultimately foreseen. The Debye length computed using the estimates above yields values between \SI{1}{\centi\meter} and \SI{3}{\centi\meter}. These calculations suggest that the Debye length is somewhat smaller than the size of the system (\SIrange{5}{10}{\centi\meter}). It was not the primary goal of these experiments to create an electron plasma, but rather to establish a space charge and electron density comparable to the target for pair plasma studies. Nevertheless, we have created an electron cloud that marginally qualifies as being a non--neutral plasma.

\section{Injection of Positrons Without an Electron Space Charge}

\begin{figure}
	\centering
	\includegraphics [width=0.49\textwidth]{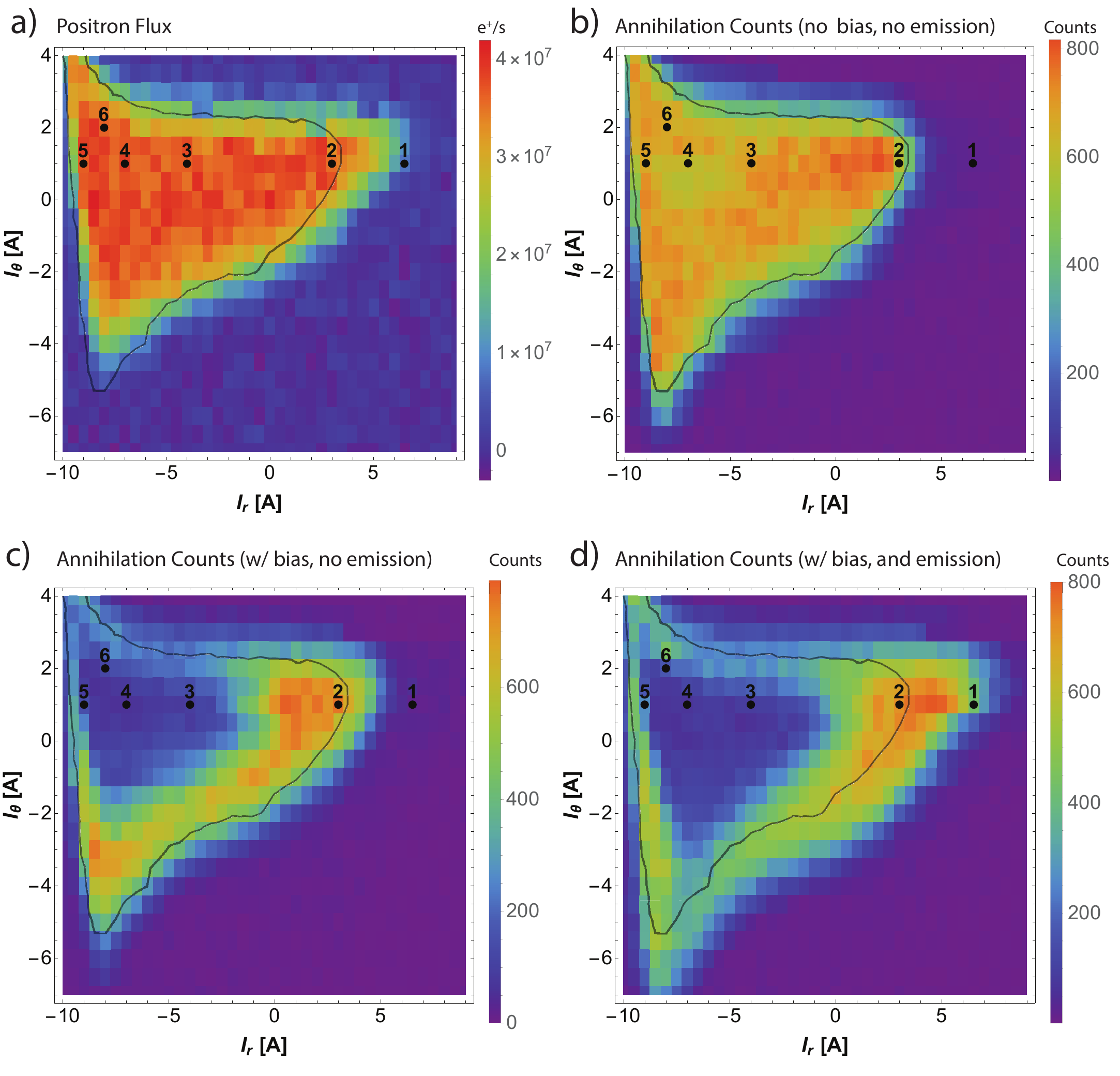}
	\caption{\si{I_r} vs \si{I_\theta} maps with color-coded positron data. The black line represents the 450 count (in \SI{1}{\second}) contour line extracted from b). The numbered points are parameter values where current profiles are measured, as explained in the text:
	a) positron current (\si{\raiseto{+}\elementarycharge\per\second}) measured at the target probe by a charge integrator,
b) annihilation counts (in \SI{1}{\second}) originating from the filament assembly measured by detector A with the filament unheated and unbiased.
c) annihilation counts with the unheated but biased (\SI{-60}[V_{acc}=]{\volt}) filament. 
d) annihilation counts for the heated \SI{4.0}[I_{fil}=]{\ampere} and biased filament. The area of injection is extended towards positive \si{I_r} values, meaning that the electron cloud allows a successful injection of positrons from parameters that are not feasible otherwise, while preserving the injection efficiency. }
	\label{fig:combined plots}
\end{figure}

The initial position of the positron beam prior to the E$\times$B drift injection is important for achieving lossless injection. To optimize this, the currents applied to the two pairs of steering coils were scanned, while the positron current was measured on the target probe, which was fully inserted at the \ang{90} position. This was done without the electron space charge. The current applied to one set of coils (\si{I_r}) moves the beam in the radial direction, and other moves the beam in the azimuthal direction (\si{I_\theta}), in the cylindrical coordinate system centered at and coaxial with the magnet. Figure \ref{fig:combined plots} a shows \si{I_r} on the horizontal axis, \si{I_\theta} on the vertical axis and the injected positron intensity with a color scale. A similar measurement was made by retracting the \ang{90} target probe and recording the annihilation radiation originating from positrons hitting the filament assembly (Figure \ref{fig:combined plots}  b) using two gamma detectors (A and B) with collimated fields of view. This set of data was acquired when the E$\times$B plates were set to \SI{ +- 150}[V_{E \times B}=]{\volt}, \SI{12}[V_{Top 1}=+]{\volt}, \SI{10}[V_{sec 1}=+]{\volt} and \SI{8}[V_{mag}=+]{\volt}, and a total counting time of \SI{1}{\second}. The remaining electrodes were grounded. It has been shown that these potentials provide lossless injection of the \SI{5}{\electronvolt} positron beam if the steering coils have been adequately adjusted \cite{Stenson_PRL_2018}. Data from detector A is shown in Figure \ref{fig:combined plots} b-d.  The counts observed by detector B are very similar, but the count rate is lower due to narrower collimation and the smaller size of the BGO crystal.

Figure \ref{fig:combined plots} a) demonstrates that, over a large parameter space, the injection efficiency is homogeneous within the accuracy of the measurements. The differences between \ref{fig:combined plots}a) and \ref{fig:combined plots}b) can be explained by the angular offset between the current probe (\ang{90}) and the field of view of the detector A (\ang{180}). A large positive \si{I_r} value corresponds to injection orbits close to the outer wall (Point 1). Such injection radii likely result in some positrons drifting out to annihilate on the outer wall before reaching the \ang{180} position \cite{Stenson_PRL_2018,Nissl_POP_2020}. Consequently, two effects can occur. First, these particles cannot be measured if they annihilate outside the detector's field of view or second, even if they are within the field of view, the count rate is lower due to the larger distance to the detector.

\begin{figure}
	\centering
	\includegraphics[width=0.49\textwidth]{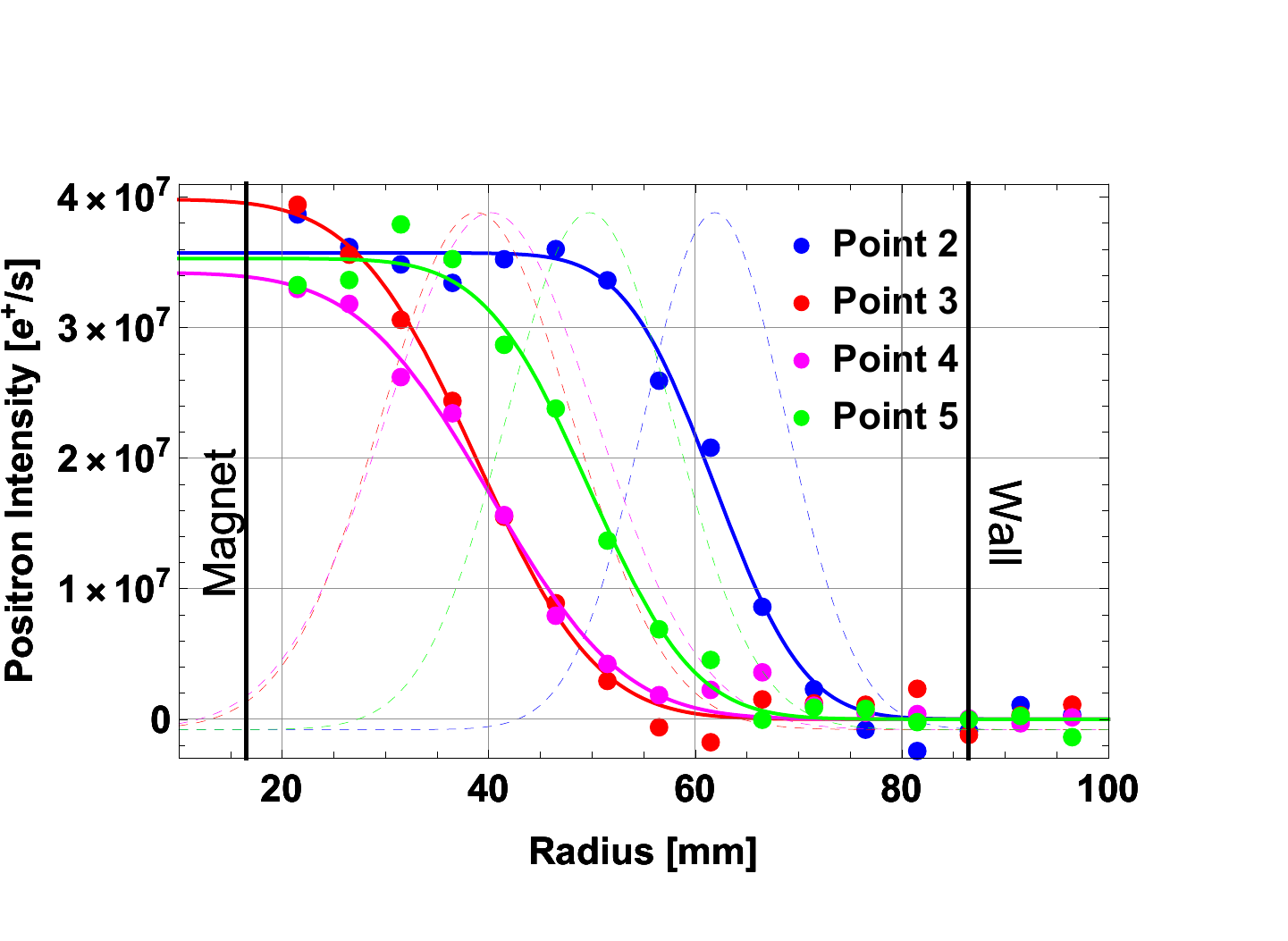}
	\caption{Injected positron intensity profiles measured at various parameters (points shown in figure \ref{fig:combined plots}) for \SI{150}[V_{E \times B}=]{\volt}. The surface of the magnet is positioned at \SI{16}{\milli\meter}. A low \si{I_r} value results in positron profiles close to the magnet. The dots show the measured values, fitted with error function (lines) and the corresponding derivative (dashed lines). These gaussians illustrate the radial distribution of the injected positrons.}
	\label{fig:current profiles points_150}
\end{figure}

While this type of measurement illustrates the total intensity of positrons for different injection conditions, it does not provide information about the radial position of trapped positrons. Radial profiles are measured by step-wise retracting the target probe at \ang{90} while measuring the positron intensity (Figure \ref{fig:current profiles points_150}). Consider several profiles with the same azimuthal steering coil current \SI{1.0}[I_{\theta}=]{\ampere}. Changing the radial steering current \si{I_r} from negative to positive values, the positron beam moves onto orbits that are closer or farther away from the magnet. The profiles shown in Figure \ref{fig:current profiles points_150} illustrate this: at Point 3 positrons are injected onto orbits closest to the magnet, the profile is shifted gradually towards the midradius for point 2. Point 1 does not show a measurable positron current.

\section{Injection of Positrons into an Electron Space Charge}
	
When establishing the electron space potential, two distinctive effects occur. One is the effect of the additional potential created by the negatively biased filament, while the other is the effect of the electron cloud. It is important to note that the rate of annihilation between the injected positrons and the electrons making up the space charge is expected to be entirely negligible in the parameter regime of these and future experiments \cite{stoneking_2020}.

\subsection{Effect of the Negative Filament Bias}

\begin{figure}
	\centering
	\includegraphics[width=0.5\textwidth]{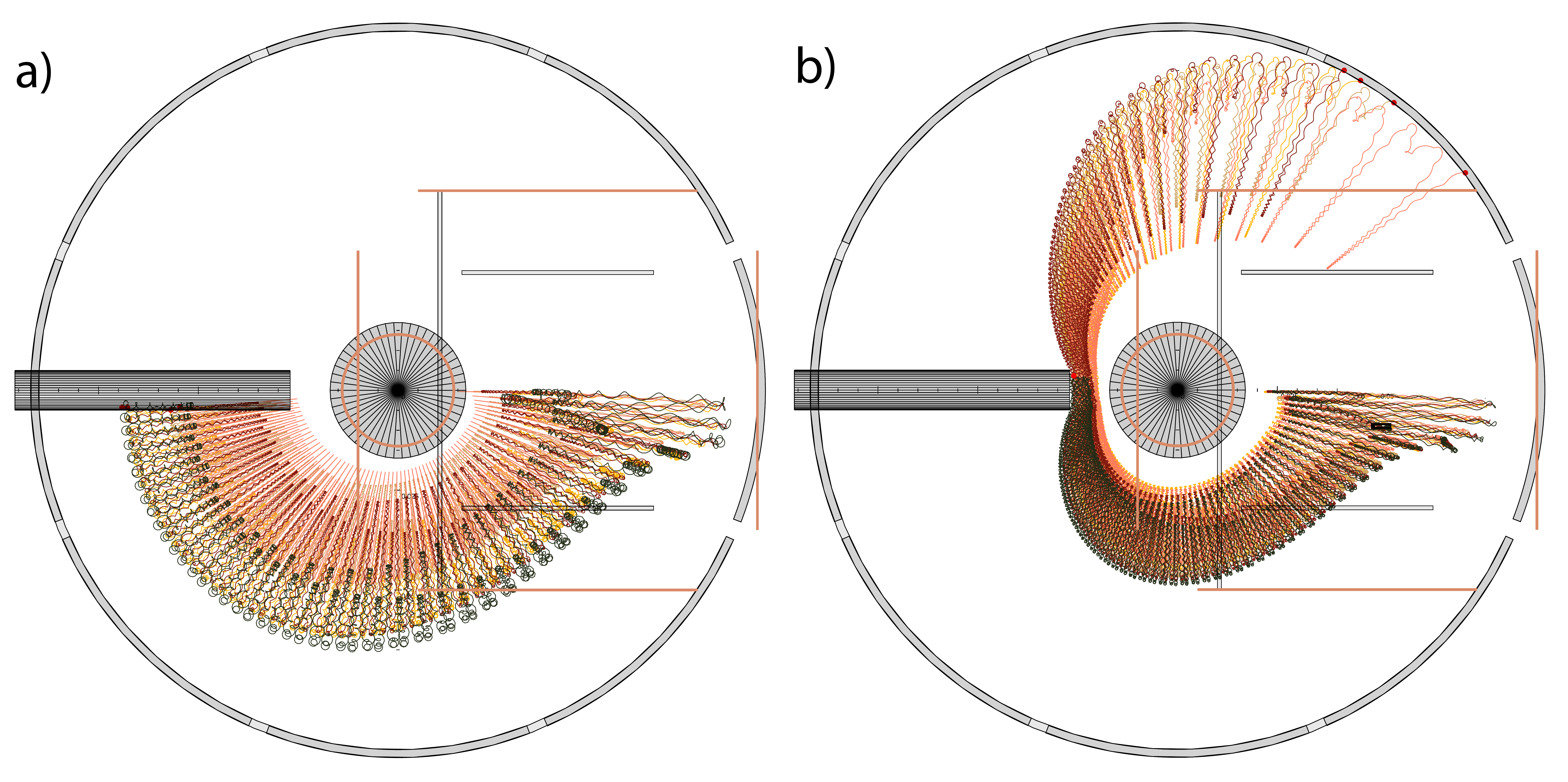}
	\caption{Simulations of injected positron trajectories. a) filament assembly grounded b) filament assembly biased to \SI{-60}{\volt}. Positrons pass through the gap between magnet and filament when it is biased.}
	\label{fig:simulation}
\end{figure}

The effect of the filament potential can be seen when comparing Figures \ref{fig:combined plots} b) and \ref{fig:combined plots} c). By applying the negative potential, without heating the filament, the formerly uniform parameter regions of efficient injection shows a hollow feature. Particle trajectory simulations \cite{Nissl_POP_2020} indicate that the additional potential applied to the filament of \SI{-60}{\volt} compresses the positron orbits towards the magnet, allowing a large fraction to pass through the \SI{1}{\centi\meter} gap between the filament assembly and the magnet. Consider points 3 and 4 in Figures \ref{fig:combined plots} and \ref{fig:current profiles points_150}. When positrons are injected with these steering coil current settings, their orbits are closest to the magnet (Figure \ref{fig:current profiles points_150}) and are located inside the hollow feature in Figure \ref{fig:combined plots} c). Figure \ref{fig:simulation} depicts particle simulations for the experimental injection scenario where the positrons are injected at radial positions close to the magnet, with the filament assembly biased to \SI{0}{\volt} and \SI{-60}{\volt}, respectively. When grounded, positrons annihilate on the filament assembly. When biased negatively, the E$\times$B drift moves the beam inwards, allowing some of the positrons to pass it \cite{Nissl_POP_2020}. As these positrons continue to toroidally drift around the magnet, they leave the collimated fields of view of the two gamma detectors and annihilate at an unobserved position.

\begin{figure}
	\centering
	\includegraphics[width=0.49\textwidth]{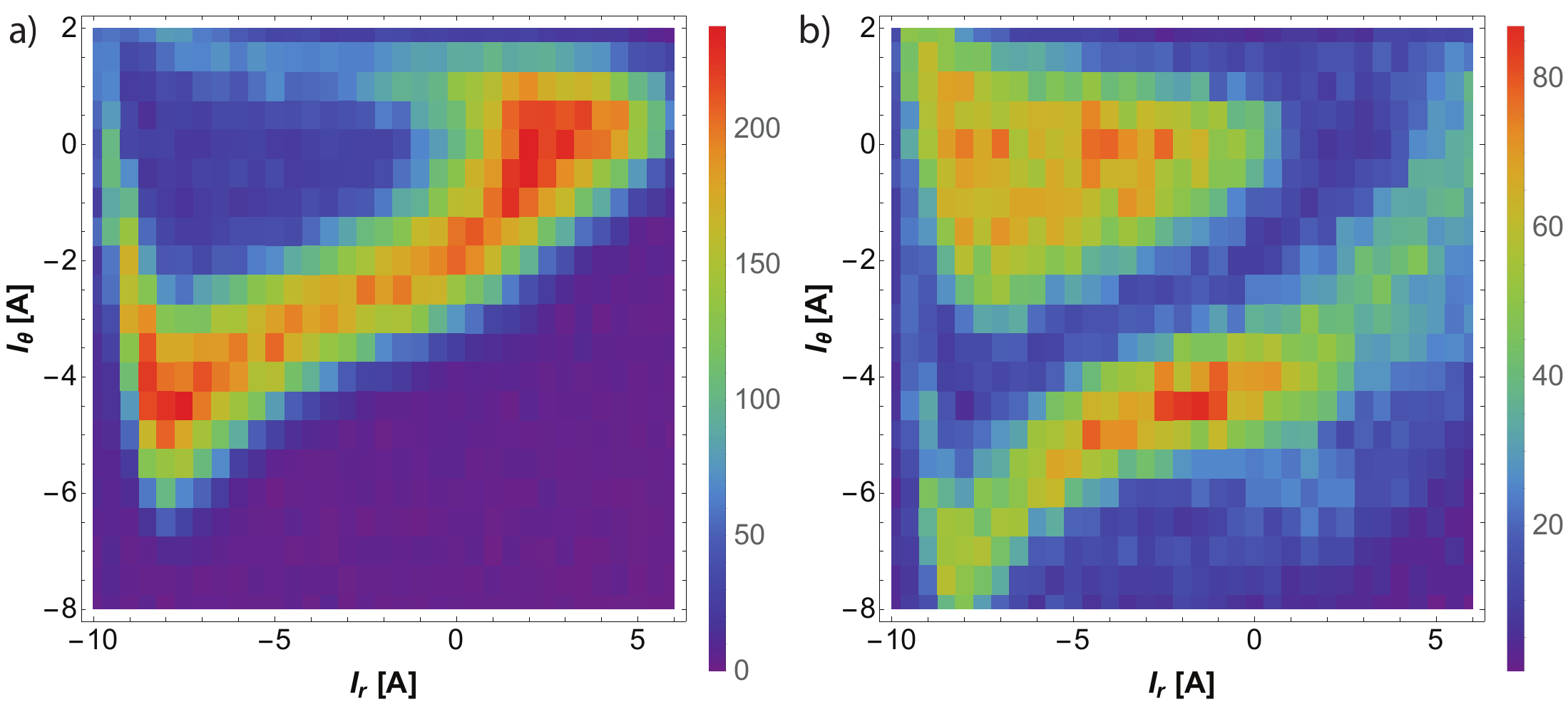}
	\caption{Annihilation data acquired in a similar campaign with lower positron intensity. a) detector A. The "hollow" area in the top left corner is also visible here (\SI{-60}[V_{acc}=]{\volt}, \SI{0}[I_{fil}=]{\ampere}, \SI{150}[V_{E \times B}=]{\volt}). b) detector B: The histogram appears like an inverted version, indicating that the injection process is not halted by the negative filament, but rather positrons annihilate at different locations (\SI{-60}[V_{acc}=]{\volt}, \SI{0}[I_{fil}=]{\ampere}, \SI{150}[V_{E \times B}=]{\volt}).}
	\label{fig:1 beamtime}
\end{figure}

Further support for this explanation for the hollow feature comes from a similar experimental campaign, during which a third detector C was installed on the equatorial plane of the magnet behind wall sector 8 aimed at observing gammas from positrons annihilating on the target probe at \ang{90}, when it was inserted. Due to space constraints, however, detector C's field of view was wider than that for detector A and also included the magnet and space charge filament (Figure \ref{fig:1 beamtime} a. The field of view of this detector included the magnet, and the sector 5 filament. The positron intensity during this campaign was lower (\SI{1.5 e7}{\raiseto{+}\elementarycharge\per\second}) and the counting time was \SI{1}{\second}.
Figure \ref{fig:1 beamtime} compares the annihilation counts from the two detectors (A and C), acquired simultaneously while \si{I_R} and \si{I_{\theta}} were scanned. The data from detector A (Fig. \ref{fig:1 beamtime}a) again shows the hollow region when the space charge filament is biased (but not emitting). The data from detector C, however, shows an approximately inverted pattern; the highest count rates at this detector correspond to regions of suppressed counts at detector A. This means that the injection process for positrons remains efficient when the filament is biased negatively, but for positron orbits close to the magnet, annihilation takes place outside the field of view of detector A and most likely on the magnet or near the E$\times$B plates, where they are observed with detector C.

\subsection{Effect of the Electron Space Charge}

When applying the heating current of \SI{4.0}[I_{fil}=]{\ampere} to the electron source (Figure \ref{fig:combined plots} d), the range of steering coil currents resulting in successful injection and \ang{180} transit around the magnet is extended towards positive \si{I_r} values. Further, the maximum number of annihilation counts observed is not altered, indicating that even though the two additional potentials created by the bias on the filament and by the space charge are significantly higher than the positron beam energy, the injection efficiency remains nearly lossless. This finding is of major importance for the future APEX project, as it demonstrates experimentally that the presence of an electron cloud affects but does not prohibit the injection of positrons; if anything, it improves it for some injection conditions.

To further verify that this effect truly originates from the presence of electrons, a measurement investigating the correlation between the emission current of the filament and the number of registered annihilation events on the filament probe was conducted. The points 1 and 6 (Figure \ref{fig:combined plots} a,b), are selected as steering coil current settings. The filament temperature threshold for electron emission is reached for a heating current above \SI{3.0}[I_{fil}=]{A}; even though the filament is biased and heated, significant electron emission does not take place for heating currents less than this value. Increasing the heating current from \SI{3.0}{\ampere} to \SI{4.0}{\ampere} in steps of \SI{0.05}{\ampere} increases the emission current from \SI{0.0}{\milli\ampere} to \SI{0.7}{\milli\ampere}. By measuring the correlation between the emission current (which establishes the space charge) and the annihilation signal from the filament probe, we can largely separate the effects of the bias on the filament (which changes very little as the heating current varies from \SIrange{3.0}{4.0}{\ampere}) from the effects due to the presence of the space charge.
	
\begin{figure}[htb]
	
	\centering
	
	\begin{minipage}[t]{0.5\textwidth}
		\centering
		\includegraphics[width=0.7\textwidth]{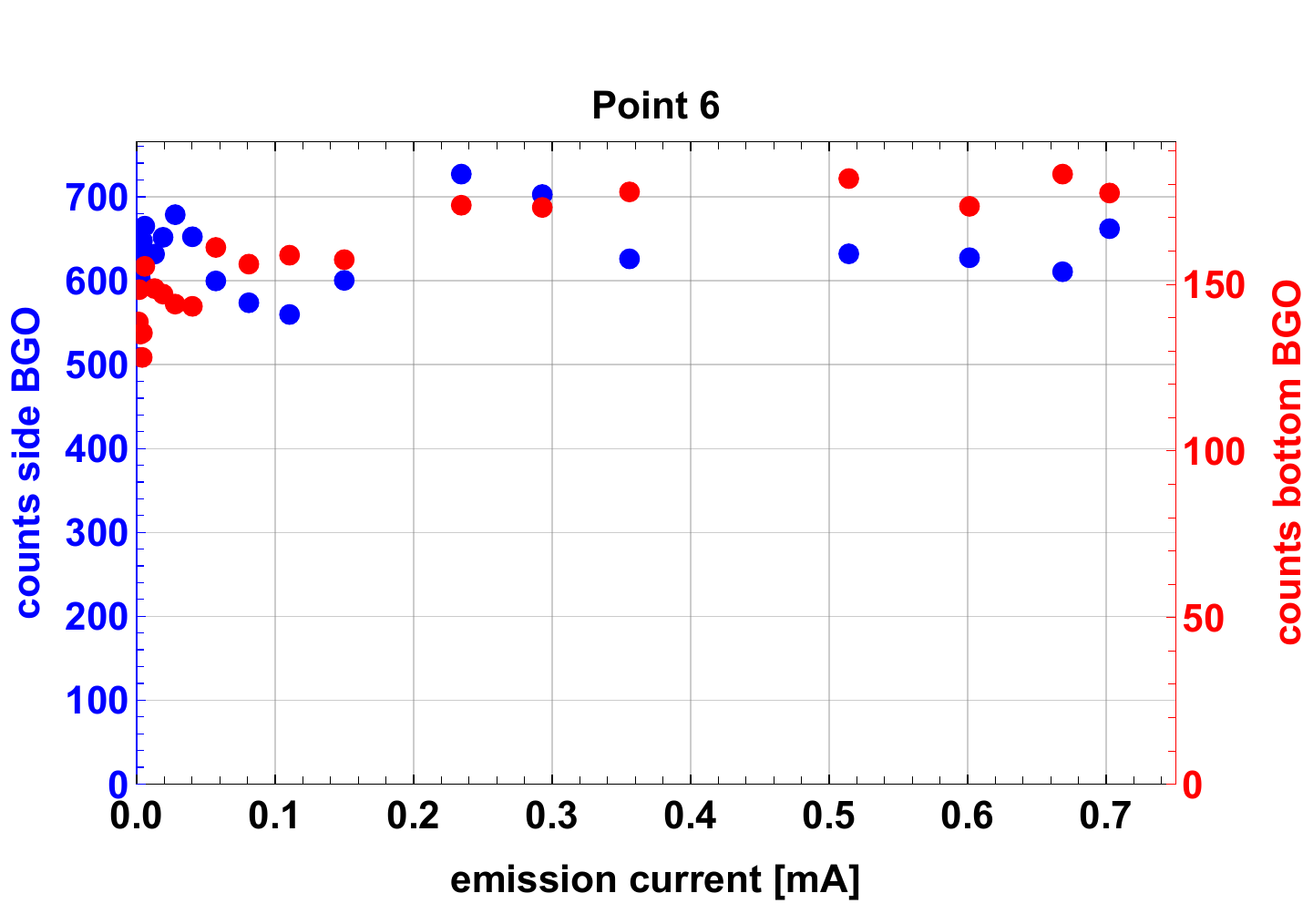}
		\caption{Annihilation count rate as function of emission current at Point 6 (in Figure \ref{fig:combined plots}a), \SI{150}[V_{E \times B}=]{\volt} (detector A (blue) and detector B (red)). The registered counts remain roughly constant, as injection for this steering current configuration is not affected by the electron cloud.}
		\label{Fig:current_counts_150_6}
	\end{minipage}
	\hfill
	\begin{minipage}[t]{0.5\textwidth}
		\centering
		\includegraphics[width=0.7\textwidth]{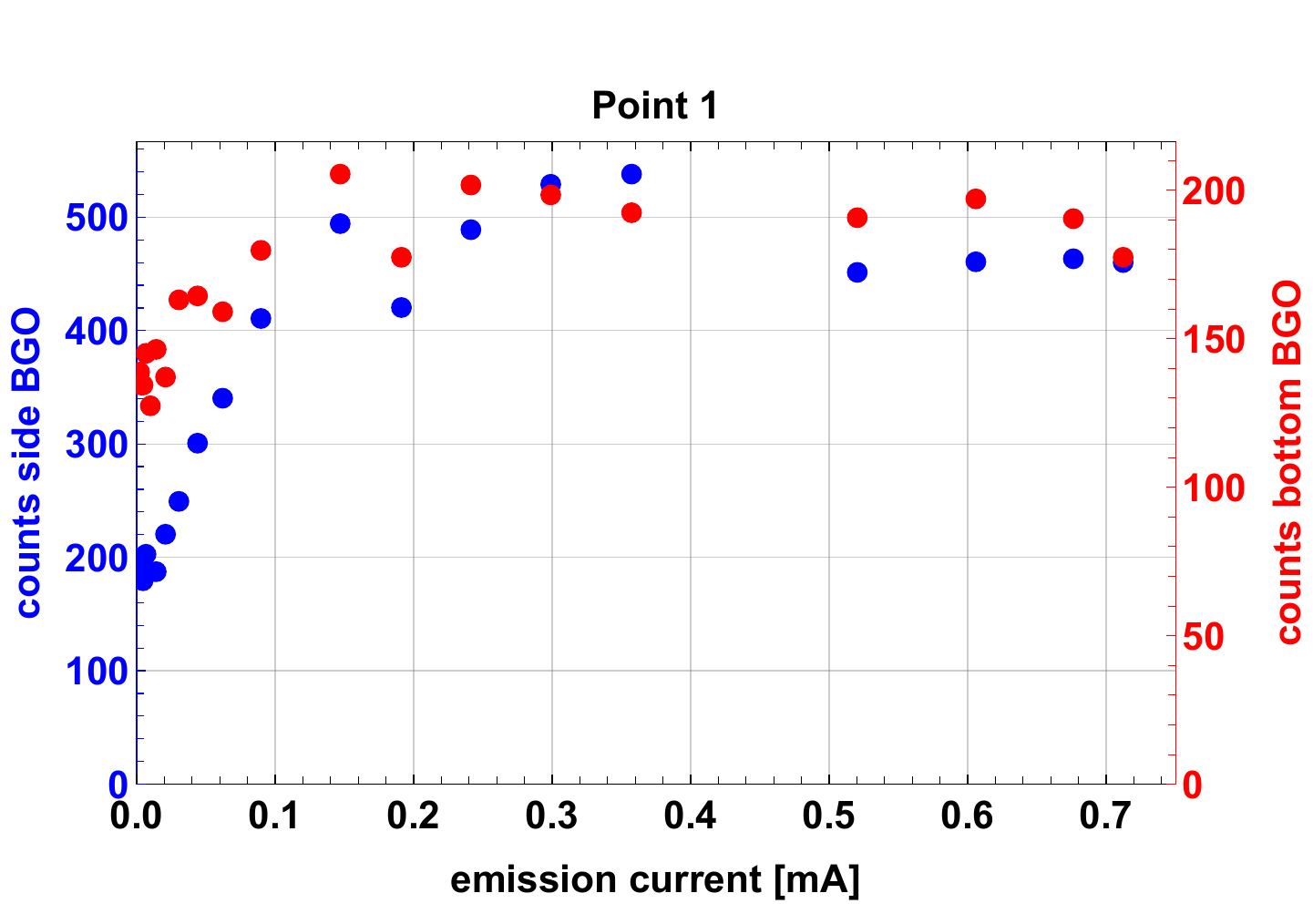}
		\caption{Annihilation count rate as function of emission current at Point 1, \SI{150}[V_{E \times B}=]{\volt}. \SI{150}[V_{E \times B}=]{\volt} (detector A (blue) and detector B (red)). The rising countrate shows that the electron cloud enhances the injection efficiency for these injection parameters.}
		\label{Fig:current_counts_150_1}
	\end{minipage}

\end{figure}

Most injection parameter settings are not affected by the presence of the electron cloud (inside the black contour in Figure \ref{fig:combined plots} d)) such as Point 6, where the count rates remain almost constant as the emission current is increased (Figure \ref{Fig:current_counts_150_6}). By contrast, without an electron cloud, positron injection is ineffective for the steering coil settings represented by Point 1. Increasing the emission current and therefore the negative space potential increases the annihilation signal (Figure \ref{Fig:current_counts_150_1}). This effect saturates for emission currents larger than \SI{0.15}[I_{emission}=]{\milli\ampere}, corresponding to a filament heating current of \SI{3.65}[I_{fil}=]{\ampere} and a maximum space potential of approximately \SI{-20}{\volt} (Figure \ref{Fig:negative_outside}).

\section{Conclusions}
	
The prototype permanent magnet trap was used to demonstrate on of the techniques required to generate an electron--positron pair plasma in a levitated dipole trap in the near future. An important step towards this goal is the injection of low--energy positrons into an electron cloud. The magnetic drifts of the two species propagate in opposite directions around the magnet. As their starting points in the trap are \ang{180} apart from each other, their orbits overlap, and the influence of the electron cloud on the injection of positrons is investigated. The thermionic electron source generates a negative space potential up to  \SI{-42}{\volt}. The electrostatic potentials, produced by the bias voltage on the electron source and the space charge exceed the longitudinal positron energy and the remaining electrostatic potentials in the trap significantly. Nevertheless, 100\% injection efficiency is preserved. What initially appeared to be a 'hole' in the region of parameter space for lossless injection is instead found to be caused by deeply injected positrons that drift between the filament assembly (doubling as an annihilation target) and the magnet, then annihilating outside the field of view of the detectors. Further, the parameter range of efficient positron injection is extended noticeably. The method used here for creating a dense electron cloud is not compatible with our ultimate plans for creating electron-positron plasma.  At this time, we are considering at least two approaches which will be tested in future experiments: 1) injecting electrons from a source placed at the edge of the confinement region as in Saitoh, \emph{et al.} \cite{Saitoh_2010}, and 2) injection from an upstream source that brings electrons into the trap along a path similar to that of the positrons.

In summary, we have demonstrated successful injection of positrons in the prototype dipole trap in the presence of an electron space charge. The measured electron density is comparable to the target density of the APEX electron--positron pair plasma. Such a result was by no means assured, given the E$\times$B drift technique we employ to inject positrons.  It might have been the case that the electron plasma shielded the fringing fields from the E$\times$B plates in a way that prevented positron injection. The results therefore represent an important validation of our approach to achieving a pair plasma.


\section{Acknowledgments}

The measurements described in this work were performed at the NEPOMUC positron beam facility installed at the FRM II at the Heinz-Maier-Leibnitz Zentrum (MLZ), Garching, Germany. Funding was received from the European Research Council (ERC-2016-ADG no. 741322), the Deutsche Forschungs Gesellschaft (Hu 978/15, Hu 978/16, Sa 2788/2), the Max-Planck-Institute for Plasma Physics (IPP), the Helmholtz Association and the UCSD Foundation.

\section{Data Availability}
The data that support the findings of this study are available from the corresponding author upon reasonable request.

\section{References}
\bibliography{aipsamp}

\end{document}